\documentclass[
    ,final            
  ]
  {aipproc}

\layoutstyle{6x9}

\usepackage{bm}

\begin{document}

\title{Scale dependent alignment between velocity and magnetic field fluctuations 
in the solar wind and comparisons to Boldyrev's phenomenological theory}


\author{J.~J.~Podesta}{
  address={Space Science Center, University of New Hampshire, Durham, NH},
  email = {jpodesta@solar.stanford.edu}
}

\author{A. Bhattacharjee}{
  address={Space Science Center, University of New Hampshire, Durham, NH}
}

\author{B. D. G. Chandran}{
  address={Space Science Center, University of New Hampshire, Durham, NH}
}

\author{M.~L.~Goldstein}{
  address={NASA Goddard Space Flight Center, Greenbelt, MD}
}

\author{D.~A.~Roberts}{
  address={NASA Goddard Space Flight Center, Greenbelt, MD}
}

\begin{abstract}
A theory of incompressible MHD turbulence recently developed by Boldyrev predicts 
the existence of a scale dependent angle of alignment between velocity and magnetic 
field fluctuations that is proportional to the lengthscale of the fluctuations to 
the power 1/4. 
In this study, plasma and magnetic field data from the {\it Wind} spacecraft are used to 
investigate the angle between velocity and magnetic field fluctuations in the solar wind 
as a function of the timescale of the fluctuations and to look for the power law scaling 
predicted by Boldyrev.  Because errors in the velocity vector can create 
large errors in the angle measurements, particularly at small scales,  the 
angle measurements are suspected to be unreliable except at the largest inertial 
range scales.  For the data at large scales the observed power 
law exponents range from 0.25 to 0.34, which are somewhat larger than Boldyrev's 
prediction of 0.25.  The results suggest that the angle may scale 
like a power law in the solar wind, at least at the largest inertial range scales, 
but the observed power law exponents appear to differ from Boldyrev's theory.   

\end{abstract}

\maketitle


\section{Introduction}

\indent \indent
Phenomenological turbulence theories developed independently by Iroshnikov 
\cite{Iroshnikov:1964} and Kraichnan \cite{Kraichnan:1965} extended and adapted the 
ideas of Kolmogorv's well known theory of hydrodynamic turbulence to incompressible 
MHD turbulence.   Both Iroshnikov and Kraichnan predicted an equipartition of energy 
between kinetic and magnetic field fluctuations in the inertial range and an 
energy spectrum proportional 
to $k^{-3/2}$.  But these early studies neglected the anisotropy
of the turbulence which was subsequently found to be ubiquitous in both
laboratory plasma experiments and in theoretical studies based on analysis and
simulations of the equations of resistive incompressible MHD \cite{Shebalin:1983}.
Turbulence in magnetized plasmas is spatially anisotropic with
the local mean magnetic field providing a natural axis of symmetry.

More recent theories of incompressible MHD turbulence incorporate the anisotropy of 
the turbulence into the theory in a fundamental way.  An influential 
theory of this kind is the theory of Goldreich and Sridhar (1995) \cite{GS:1995},
hereafter GS95, with important corrections by Goldreich and Sridhar (1997) \cite{GS:1997}.  
GS95 introduced the idea of `critical balance' in which there is a 
balance between the eddy turnover time, or energy cascade time,
and the Alfv\'en crossing time of two wavepackets propagating in opposite
directions along the local mean magnetic field.  As a consequence of critical
balance, the GS95 theory
predicts that for a typical wavepacket the wavelengths parallel and perpendicular
to the local field satisfy the anisotropic relation $\lambda_\perp \propto
\lambda_\parallel^{2/3}$ and the perpendicular energy spectrum is 
proportional to $k_\perp^{-5/3}$.

The decade following the publication of GS95 saw improved simulations of
incompressible MHD turbulence in two and three dimensions, many of which showed that for 
plasmas with a strong mean magnetic field $|\bm B_0|$ comparable to or greater than the 
r.m.s.~magnetic field fluctuations the perpendicular energy spectrum scales like 
$k_\perp^{-3/2}$ in contradiction to the GS95 theory \cite{Maron:2001, Ng:2003, 
Muller:2003, Muller:2005}.  To resolve this discrepancy Boldyrev  
\cite{Boldyrev:2005, Boldyrev:2006} developed a new phenomenological theory with
a critical balance condition different from that of GS95. 
A somewhat different theory was derived by Beresnyak and Lazarian \cite{Beresnyak:2006}.
The key new idea introduced by Boldyrev is that as energy cascades from large
to small scales through the inertial range the velocity and 
magnetic field fluctuations undergo an alignment process whereby the average
angle $\theta$ between $\delta \bm v_\perp$ and $\delta \bm b_\perp$ is a
monotonically decreasing function of scale.  In his theory,
Boldyrev predicts that the angle obeys the scaling law
$\theta \propto \lambda_\perp^{1/4}$ and that the perpendicular energy spectrum is 
proportional to $k_\perp^{-3/2}$.  

Evidence for Boldyrev's alignment process and for the scaling law
$\theta \propto \lambda_\perp^{1/4}$ have been obtained through direct numerical
simulations of forced, steady state incompressible MHD turbulence in three dimensions
by Mason et al. \cite{Mason:2006, Mason:2008}.  These simulations 
demonstrate that for a number of different types of forcing functions
the system develops an inertial range spanning approximately
one decade in wavenumber where the perpendicular energy spectrum is 
proportional to $k_\perp^{-3/2}$ and, presumably over the same range, 
$\theta \sim \lambda_\perp^{1/4}$.  The successful demonstration of Boldyrev's 
alignment process by Mason et al. prompted us to consider
whether this alignment process may also take place in the solar wind,
a naturally occuring turbulent plasma that is directly accessible to
in-situ spacecraft measurements \cite{Marsch:1991, Goldstein:1995, 
Tu_Marsch:1995, Bruno_Carbone:2005}.
Therefore a study was undertaken to investigate the possible existence of a
scale dependent alignment between
velocity and magnetic field fluctuations in the solar wind.
The results of this study shall now be briefly summarized.

\section{The quantity being measured}

\indent \indent
Boldyrev and his colleagues measure the angle $\theta$ using the formula 
\cite{Mason:2006, Mason:2008}
\begin{equation}
\theta(r) = \sin^{-1} \left( \frac{\langle |\delta \bm v_\perp \times  
\delta \bm b_\perp|
\rangle} {\langle |\delta \bm v_\perp|\cdot |\delta \bm b_\perp| 
\rangle}\right),  \label{theta_boldyrev}
\end{equation}
where $\delta \bm v_\perp$ and $\delta \bm b_\perp$ are the projections of the 
fluctuations $\delta \bm v=\bm v(\bm x+\bm r)-\bm v(\bm x)$ and 
$\delta \bm b=\bm b(\bm x+\bm r)-\bm b(\bm x)$
onto the plane perpendicular to the {\it local} mean magnetic field $\bm B_0(\bm x)$,
respectively, and the displacement $\bm r$ is perpendicular to $\bm B_0(\bm x)$.
It is desirable but not possible
to employ precisely the same formula (1) when analyzing solar wind data.  
A single spacecraft essentially performs measurements along the solar wind
flow direction. 
Because the flow is super-Alfv\'enic,
Taylor's ``frozen turbulence'' hypothesis may be used to relate the time  $\tau$ 
between measurements to a spatial separation $r=V_{\rm sw}\tau$ along the average 
flow direction, approximately the radial direction in heliocentric coordinates.  
The angle between the mean magnetic field and the average flow direction varies 
significantly due to variations in the direction of the mean magnetic field, 
but it is often near 45 degrees, the inclination
of the Parker spiral at 1 AU.  Even though the theory is formulated for fluctuations
between two points $\bm x$ and $\bm x+\bm r$ inclined at 90 degrees to the local
mean magnetic field, we expect that any alignment (if it exists) will still be
measureable at inclinations of 45 degrees \cite{Podesta:2008}.

The solar wind data employed in this study consists of simultaneous measurements of the 
average velocity vector and magnetic field vector from instruments on the {\it Wind} 
spacecraft.  The data was acquired near the orbit of the earth at 1 AU
when {\it Wind} was far away from the influences of the Earth's 
magnetosphere and bow shock.  The data have an approximate 24-second cadence.  
The intervals analyzed have durations from several days to several months.
For a given time scale $\tau$ the fluctuations $\delta \bm v=\bm v(t+\tau)-\bm v(t)$ and 
$\delta \bm b=\bm b(t+\tau)-\bm b(t)$ are projected 
onto the plane perpendicular to the {\it local} mean magnetic field $\bm B_0(t)$
to obtain $\delta \bm v_\perp$ and $\delta \bm b_\perp$, respectively.  
The average angle $\theta(\tau)$ is then computed using equation (1).  The local
mean magnetic field $\bm B_0(t)$ is scale dependent and is defined as the average 
(vector) over the interval from $t-\tau$ to $t+2\tau$.  The essential 
difference between this study and the method employed by Boldyrev is that 
the displacement $\bm r$ is not perpendicular to $\bm B_0$ in the solar wind.

\section{Results}

\indent \indent
An example of the  angle measurements obtained using {\it Wind} data 
is shown in Figure 1.  
\begin{figure}
  \includegraphics[width=.5\textwidth]{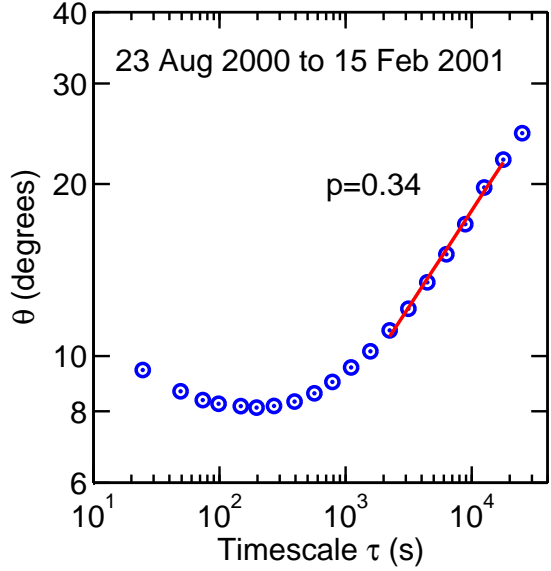}
  \caption{The average angle $\theta$ as a function of the timescale $\tau$ (circles) obtained
from {\it Wind} 3DP data. The solid red line is the best fit line over the range from 
$2\times 10^3$ to $2\times 10^4$ s.  The slope of the line yields the power law exponent
$p=0.34$.  Each data point represents an average over many measurements $(\sim 4\times 10^5)$.}
\end{figure}
Note that the measurements span most of 
the inertial range, the range of timescales from approximately 3 s to $10^4$ s.
All the data analyzed in this study show an approximate power law scaling
at the largest inertial range scales (a straight line on a log-log plot)
followed by a roll-over of the curve at medium to small scales.  This roll-over
is clearly a deviation from the power law behavior seen at the largest scales.
We suspect that the roll-over is due to small errors in the plasma velocity measurements
which can cause large errors in the angle measurements at small scales.
For this reason, estimates of the power law exponent $p$ in the scaling law 
$\theta(\tau) \propto \tau^p$ are only performed at the largest inertial range
scales from $2\times 10^3$ to $2\times 10^4$ s.  A linear least-squares fit
on a log-log plot yields the best fit line and power law exponent $p=0.34$ for the data 
shown in Figure 1.

Proton velocity moments obtained using the 3DP instrument on {\it Wind} are 
subject to measurement errors, like any measurement apparatus.  Let $\varepsilon$
denote the typical measurement error for the velocity components $v_x$, $v_y$, and $v_z$.
Then, the typical error for the velocity difference $\delta v_x$ is $2\varepsilon$.
Suppose the errors in the magnetic field $\delta \bm b$ are negligible compared to
the errors in the velocity $\delta \bm v$.  To analyze the error, choose
the $z$-axis parallel to the vector $\delta \bm b$ and let the vector $\delta \bm v$ 
lie in the $xz$-plane.  Then an error $2\varepsilon$ in the measured velocity $\delta v_x$ 
will create an error $\delta \theta$ in the angle $\theta$ as determined by the equations
\begin{equation}
\sin(\theta)=\frac{\delta v_x}{|\delta \bm v|},\qquad 
\sin(\theta +\delta \theta)=\frac{\delta v_x\pm 2\varepsilon}{|\delta \bm v|}.
\end{equation}
Assuming the angles are all small, it follows that $\delta \theta \simeq 2
\varepsilon/|\delta \bm v|$. If $|\delta \bm v|$ is replaced by its r.m.s.~amplitude 
$\delta v$, then the quantity $\delta \theta = 2 \varepsilon/\delta v$ can be 
thought of as the smallest resolvable change in angle.

\begin{figure}
  \includegraphics[width=.5\textwidth]{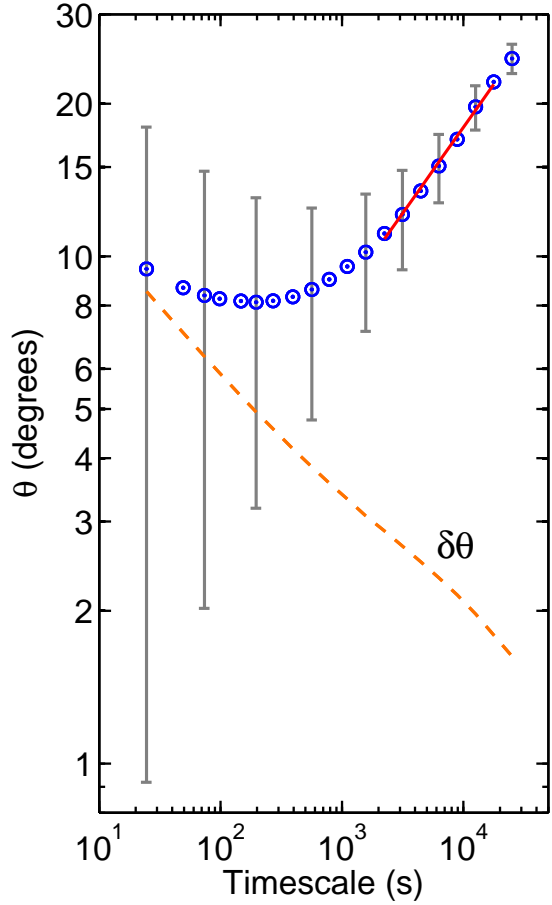}
  \caption{The average angle $\theta$ (circles), the error $\delta \theta 
= 2 \varepsilon/\delta v$ assuming $\varepsilon=0.76$ km/s (dashed),
and the corresponding error bars $\theta\pm \delta \theta$ for the same data 
shown in Figure 1. Note that these are not true error
bars, only a rough indication of the smallest resolvable angle at each scale.}
\end{figure}

This rough error estimate can be used to derive rough error bars
for the angle measurements in Figure 1.  We emphasize that these are not true error
bars, only a rough indication of the smallest resolvable angle at each scale.
Assuming a typical value $\varepsilon=0.76$
km/s, the rms velocity fluctuations $\delta v(\tau)$ are computed from the data
and used to compute the estimated error $\delta \theta = 2 \varepsilon/\delta v$.
The result for $\delta \theta$ is shown by the dashed orange line in Figure 2.
Note that at the smallest scales the magnitude of $\delta \theta$ is comparable to 
the measured average angle $\theta$.  This leads us to believe that the measurements at the smallest scales are 
unreliable.  The error bars in Figure 2 furthermore suggest that data in the middle
of the range are unreliable too.  Only the data at the largest inertial range scales 
are considered to be sufficiently reliable for the purpose of estimating
the power law exponent in the scaling law.

Results for the power law exponents estimated from linear least-squares fits over the 
range of timescales
from $2\times 10^3$ to $2\times 10^4$ s are shown in Table 1.  On average, the values
for the power law exponent $p$ obtained from the data are larger than the value 0.25 
predicted by Boldyrev's theory.  Therefore, the measurements are not in overall 
agreement with the theory.  However, in two of four cases studied, intervals 2 and 3, 
the values $p=0.25$ and $p=0.27$ are within 8\% of the predicted value 0.25.  
This prompts us to ask whether there may be differences in the solar wind conditions for 
intervals 2 and 3 compared to intervals 1 and 4, differences that may make it more 
likely to find agreement with the theory in the former case than in the latter.
One difference is that intervals 2 and 3 consist of nearly homogeneous low-speed wind,
devoid of any high speed streams, while intervals 1 and 4 both contain a notable presence 
of high speed streams.
Recent work by Dasso et al. \cite{Dasso:2005} has suggested that low-speed wind appears to 
be dominated by wavevectors perpendicular to the local mean magnetic field, as expected
for quasi-2D turbulence, whereas high-speed wind appears to 
be dominated by wavevectors parallel to the mean field.  If the turbulence in 
intervals 1 and 4 is not predominantly quasi-2D turbulence, then  Boldyrev's scaling law
would not be expected to apply and that may explain
why we observe deviations from Boldyrev's scaling law in these cases.  This 
interesting possibility needs further study.


\begin{table}
\begin{tabular}{ccccc}
\hline
    \tablehead{1}{c}{b}{Interval}
  & \tablehead{1}{c}{b}{Begin Date}
  & \tablehead{1}{c}{b}{End Date}
  & \tablehead{1}{c}{b}{\ \ Days\ \ }
  & \tablehead{1}{c}{b}{\ \ $p$\ \ }  \\
\hline
1 & 01 Jan 1995  & 29 Jul 1995 &  209   &  0.34 \\
2 & 15 May 1996  & 16 Aug 1996 &   93   &  0.25 \\
3 & 08 Jan 1997  & 09 Jun 1997 &  152   &  0.27 \\
4 & 23 Aug 2000  & 15 Feb 2001 &  176   &  0.34 \\
\hline
\end{tabular}
\caption{The power law exponent $p$ obtained from fits of the angle 
$\theta$ over the range from $2\times 10^3$ to $2\times 10^4$ s.}
\label{table1}
\end{table}

\section{Conclusions}

\indent \indent
The study of solar wind turbulence presented here suggests that the
angle $\theta$ appears to scale like a power law at the largest inertial
range scales, but in many cases the observed power law exponent is significantly
larger than the value 1/4 predicted by Boldyrev's theory.  There is a 
clear breakdown of the scaling behavior at medium- to small-scales. 
Whether this breakdown is partly a genuine physical effect or is solely caused
by measurement errors in the velocity data is not known at the present time.
Therefore, we do not know whether the {\it apparent} power law scaling 
seen at large inertial range scales continues to hold at medium- and small-scales.  
More precise plasma measurements with errors less than 0.1 or 0.05 km/s for each velocity 
component are needed to answer this important question.  

In half of the cases studied the power law exponent was within 8\% of the 
value 0.25 predicted by Boldyrev.  In these cases the solar wind can be characterized as 
generally low-speed wind without any high-speed streams.  The
better agreement with Boldyrev's theory in these cases may be because low-speed wind is 
more quasi-2D in nature than high-speed wind as suggested by the analysis
of Dasso et al. \cite{Dasso:2005}.  This is an interesting open question.





\bibliographystyle{aipproc}   

\bibliography{bold}

\IfFileExists{\jobname.bbl}{}
 {\typeout{}
  \typeout{******************************************}
  \typeout{** Please run "bibtex \jobname" to optain}
  \typeout{** the bibliography and then re-run LaTeX}
  \typeout{** twice to fix the references!}
  \typeout{******************************************}
  \typeout{}
 }

\end{document}